\documentclass[12pt]{iopart}

\usepackage{iopams}
\usepackage{amsfonts}
\usepackage{graphicx}
\usepackage{cite}
\usepackage{color}
\usepackage{times}

\newcommand{\ket}[1]{\vert{#1}\rangle}

\newcommand{\an}[1]{{#1}^{\phantom\dag}}
\newcommand{\cre}[1]{{#1}^\dag}

\newcommand{\hc}{{\rm h.c.}}
\newcommand{\var}{\varepsilon}
\newcommand{\gen}{\mathcal{L}}
\newcommand{\ev}{\mu{\rm eV}}

\usepackage{color}
\usepackage{soul}

\definecolor{darkred}{rgb}{0.8,0.1,0.1}
\definecolor{darkblue}{rgb}{0.1,0.1,0.7}

\begin{document}

\title{Dephasing-assisted transport in linear triple quantum dots}

\author{L.~D.~Contreras-Pulido, M.~Bruderer, S.~F.~Huelga and M.~B.~Plenio}
\address{Institute of Theoretical Physics, Ulm University, D-89069 Ulm, Germany and
Center for Integrated Quantum Science and Technology, Ulm University, D-89069 Ulm, Germany}

\date{\today}

\begin{abstract}
Environmental noise usually hinders the efficiency of charge transport through coherent quantum systems; an exception is dephasing-assisted
transport (DAT). We show that linear triple quantum dots in a transport configuration and subjected to pure dephasing exhibit DAT if
the coupling to the drain reservoir exceeds a threshold. DAT occurs for arbitrarily weak dephasing and the enhancement can be directly
controlled by the coupling to the drain. Moreover, for specific settings, the enhanced current is accompanied by a reduction of the relative
shot noise. We identify the quantum Zeno effect and long-distance tunnelling as underlying dynamic processes involved in dephasing-assisted
and -suppressed transport. Our analytical results are obtained by using the density matrix formalism and the characteristic polynomial
approach to full counting statistics.
\end{abstract}


\maketitle

\section{Introduction}

The presence of environmental noise is generally considered to be an unavoidable hindrance
to efficient transport of charge or energy through quantum systems. The general view
is that transport in quantum systems relies on their coherence which is inevitably reduced by
interactions with an external noisy environment. However, recently environmental noise has
been found to play a positive role in transport. Motivated by experiments showing the presence
of quantum beating in photosynthetic systems~\cite{lee07,prokhorenko02,engel07}, subsequent
theoretical work pointed out that the efficiency of the excitation energy transfer
in light-harvesting complexes during photosynthesis can benefit from the presence of
environmental noise~\cite{mohseni08,plenio08,caruso09,huelga13}. The positive influence of dephasing on
the transport efficiency, known as dephasing-assisted transport (DAT), was identified as one of
the fundamental mechanisms involved.

For achieving a better understanding of this mechanism it is therefore very desirable to implement and explore
DAT in the laboratory under controlled conditions. Several proposals in this direction have been made, including
transport through quantum optical systems~\cite{caruso11}, information transmission in quantum information
platforms~\cite{caruso10}, and heat transport through trapped-ion crystals~\cite{bermudez13}. The problem of
dephasing-enhanced transport in homogeneous linear chains of fermions, including fluctuations~\cite{znidaric14},
has recently attracted a lot of interest. The effect of interactions~\cite{mendoza13,mendoza13b}, diagonal
disorder~\cite{znidaric13} and the interplay between both~\cite{znidaric10} has been explored based on analytical
and numerical results.

\begin{figure}[h]
\centering
\includegraphics[width=255pt]{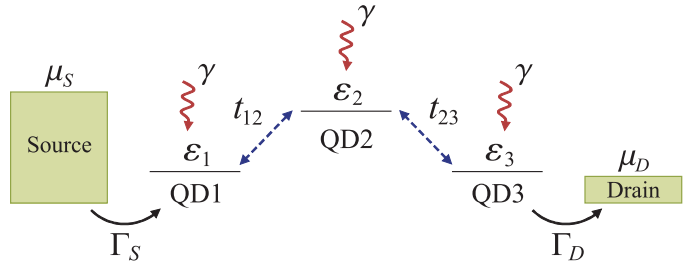}
\caption{Three quantum dots with tunnelling couplings $t_{ij}$ are connected with
coupling rates $\Gamma_S$ ($\Gamma_D$) to a source (drain) reservoir at chemical potential $\mu_S$ ($\mu_D$).
Pure dephasing with equal strength $\gamma$ acts on the single-electron
energy levels $\var_i$. The central quantum dot (QD2) is detuned from the outer dots (QD1 and QD3).
The current through the triple quantum dot is brought into the regime of dephasing-assisted transport by
increasing the coupling $\Gamma_D$ above a threshold $\Gamma_D^*$.}\label{scheme}
\end{figure}

In this work we show that the remarkable degree of control achievable in linear triple quantum dots (TQDs)
make them well suited for observing DAT and the characteristic current fluctuations associated with it. The
linear TQD is composed of tunnel-coupled single-level quantum dots (QDs), depicted schematically in figure~\ref{scheme}.
The first and last QDs are coupled to the source and drain reservoirs, respectively, and external dephasing acts
on each QD partly destroying the quantum coherence in the system. Dephasing can be due to population decay and phase
relaxation processes (pure dephasing). We here consider exclusively pure dephasing, which is not accompanied by changes of the populations.
Working with spinless electrons, we determine analytically several non-equilibrium steady-state
properties of the TQD, including the current, shot noise and the coherences of the reduced density matrix.

From  our results emerges the following picture: DAT occurs for sufficiently strong coupling to the drain and finite
detuning of the central QD. In terms of the energy levels of the TQD, dephasing-enhanced transport results from the
interplay between two mechanisms for level-broadening, namely pure dephasing and coupling to the reservoirs. From a
dynamic point of view, we argue that the equivalent of the quantum Zeno effect in quantum transport~\cite{gurvitz98,chen04}
is involved in the occurrence of DAT as an underlying process. Indeed, strong coupling to the drain
traps the coherent evolution of the charge between QD1 and QD2, thereby creating a transport bottleneck.
The positive impact of dephasing, even if of vanishingly small strength, is to impair this coherent evolution and
hence to alleviate the transport bottleneck. Therefore, DAT in the linear TQD with a single excess charge is not
due to interference effects, which in comparable systems may originate from geometry, disorder or spin degrees of
freedom~\cite{emary06,chin10,sanchez14}.

We also investigate the shot noise of the current to fully explore the existing experimental possibilities of QDs.
The problem of current fluctuations in TQDs has been addressed for triangular configurations, where the main focus
was on interference effects~\cite{groth06,poltl09,dominguez11} including the influence of a threading magnetic
field~\cite{kuzmenko06b,jiang07,emary07,delgado08a}.
For the linear TQD we find that dephasing-enhanced currents exhibits smaller relative fluctuations, i.e.,
a larger current-to-noise ratio, than the current without dephasing, at least for experimentally relevant parameter regimes.


Our proposal relies largely on the fact that the experimental study of TQDs has emerged recently, uncovering
a variety of coherent transport properties~\cite{gaudreau06,gaudreau07,gaudreau09,haug09,granger10,tarucha09b,granger12b}.
In triangular geometries, entanglement~\cite{saraga03} and effects of
interference~\cite{kuzmenko06,emary06,haug08,dominguez11,seo13,kuzmenko06b,jiang07,emary07,delgado08a,delgado08b,gaudreau09b}
have been studied, whereas in linear configurations phonon- and photon-assisted transitions as well as the influence
of spin-orbit interactions have been analyzed~\cite{vandersypen13b,jorge13}. Most relevant for our proposal, transport
measurements in a linear TQD have demonstrated superexchange, i.e., coherent long-distance tunnelling going beyond
nearest-neighbor QDs~\cite{busl13,vandersypen13,sanchez14,sanchez14b}.

Dephasing in QDs can be due to background charge noise~\cite{hayashi03,itakura03,fujisawa06},
inherent electron-phonon interactions~\cite{fedichkin04,fedichkin05,stavrou05,fujisawa06}
or capacitive coupling to nearby charge detectors, in particular the backaction of quantum
point contacts~\cite{field93,levinson97,gurvitz97,buks98,aguado00b,young10,li13,zilberberg14}.
All these interactions can be used to produce DAT provided that pure dephasing reduces the coherence
of the system while inelastic relaxation processes between electronic states are kept small.
We focus here on electron-phonon interactions as the main source of dephasing so that the dephasing
rate $\gamma$ can be controlled via the temperature of the phonon environment in the Kelvin
regime. In this case, dephasing dominates as long as the overlap between wave functions of different
local electronic states is small~\cite{itakura03,fedichkin04,fedichkin05,stavrou05}. To ensure that
environmental effects are mainly due to pure dephasing, we detune the central QD significantly,
thereby arranging the system in either a $\Lambda$-type or $V$-type configuration [cf.~figure~\ref{scheme}]
with mostly localized electronic states.  As we focus on the effect of dephasing on transport we refer to
references~\cite{fedichkin05,stavrou05,brandesrep} for microscopical aspects of electron-phonon interactions.

The paper is organized as follows: In section~\ref{model} we define the model in terms of a Lindblad master equation
that describes the dynamics of the TQD. Then in section~\ref{results} we present our results: Section~\ref{over}
contains a general overview of the transport properties of the TQD, including the exact conditions for the occurrence of DAT.
Section~\ref{coh} contains an analysis of the dephasing-induced change of the overall coherence of the TQD as well as of the
elements of the steady-state density matrix. In section~\ref{under} we discuss the physical mechanisms relevant for transport
(quantum Zeno effect and long-distance tunnelling) in detail, and in section~\ref{exp} we suggest an experimental procedure
for detecting dephasing-assisted currents. We end with the conclusions in section~\ref{conc}.

\section{Model and method}\label{model}

The linear TQD is composed of three tunnel-coupled single-level QDs, denoted by QD1, QD2 and QD3,
where QD1 and QD3 are coupled to the source and drain reservoirs, respectively [cf.~figure~\ref{scheme}].
We work in the limit of infinite bias voltage ($\mu_L\rightarrow\infty$, $\mu_R\rightarrow-\infty$).
Consequently, the Fermi functions of the reservoirs become $f_L = 1$ and $f_R = 0$ and all energy levels
$\var_i$ lie within the conduction window. The system is assumed to be
in the Coulomb blockaded regime (the intra-dot Coulomb interaction energy $U$ being the largest energy scale),
such that there is at most one extra electron in the TQD. The relevant basis states are accordingly the empty
state $\ket{0}$ and the states $\ket{1}$, $\ket{2}$ and $\ket{3}$, in which one of the dots is occupied by a
single spinless electron.

The Hamiltonian for the coherent evolution of the TQD is ($\hbar = 1$ is taken throughout the paper)
\begin{equation}\label{hdot}
H = (t_{12}\cre{d}_2\an{d}_1 + t_{23}\cre{d}_3\an{d}_2 + \hc) + \sum_i\var_i\cre{d}_i\an{d}_{i}\,,
\end{equation}
where $\var_{i}$ denotes the on-site energy of an electron on each QD and $t_{ij}$ is the tunnelling coupling
between adjacent QDs. The operators $\cre{d}_{i} (\an{d}_{i})$ create (annihilate) an electron in the
$i$th QD.
The state of the open system, including reservoirs and environment-induced dephasing, is described
by the reduced density matrix $\rho(t)$ of the TQD whose time evolution is governed by the Lindblad
master equation~\cite{breuer,rivas}
\begin{equation}\label{blad}
	\frac{\rmd}{\rmd t}\rho(t) = \gen\rho(t) = (\gen_c + \gen_S + \gen_D + \gen_\phi)\rho(t)\,.
\end{equation}
The different Lindblad operators act on $\rho(t)$ as
\begin{equation}\label{super}
\eqalign{
\gen_c\rho(t) &= -\rmi[H,\rho(t)]\,, \\
\gen_S\rho(t) &= \frac{\Gamma_S}{2}(2\cre{d}_1\rho(t)\an{d}_1 - \{\an{d}_1\cre{d}_1,\rho(t)\})\,, \\
\gen_D\rho(t) &= \frac{\Gamma_D}{2}(2\an{d}_3\rho(t)\cre{d}_3 - \{\cre{d}_3\an{d}_3,\rho(t)\})\,, \\
\gen_\phi\rho(t) &= \frac{\gamma}{2}\sum_{i}\left(2n_i\rho(t) n_i-\{n_i,\rho(t)\}\right)\,,
}
\end{equation}
where $n_i = \cre{d}_i\an{d}_i$ and $\{\;,\:\}$ stands for the anticommutator. The operator $\gen_c$
describes the coherent evolution of the system, while $\gen_S$ and $\gen_D$ model the irreversible
tunnelling of electrons out of the source and into the drain with rates $\Gamma_S$ and $\Gamma_D$,
respectively. In the infinite bias limit and within the wide-band approximation, 
these rates are constant and, moreover, the Born-Markov approximation with respect to the reservoirs
becomes exact~\cite{brandesrep,timm08}.
Finally, $\gen_\phi$ describes pure dephasing acting on all three QDs with equal strength. Dephasing causes
an exponential decay of the non-diagonal elements of $\rho(t)$ with the rate $\gamma$.

We are mainly interested in the regime of small dephasing $\gamma\lesssim t_{ij}$ for two reasons: First, values
of this order are typically observed in experiments on lateral coupled QDs~\cite{hayashi03}. Second, if the
dephasing rate $\gamma$ approaches the inter- and intra-dot Coulomb interaction, then the resulting level-broadening
leads to a substantial overlap between the single particle states $\{\ket{1},\ket{2},\ket{3}\}$ and higher energy levels
of the TQD, which are neglected here; hence we assume $\gamma\ll U$ for consistency. In addition, as shown in
experiments~\cite{hayashi03}, variations of the on-site energies $\varepsilon_i$ lead to a modification of the dephasing
rate $\gamma$; the energies $\varepsilon_i$ are therefore held constant.

\begin{figure}[t]
\centering
\raisebox{6.3cm}{{\small\bf(a)}}\includegraphics[width=170pt]{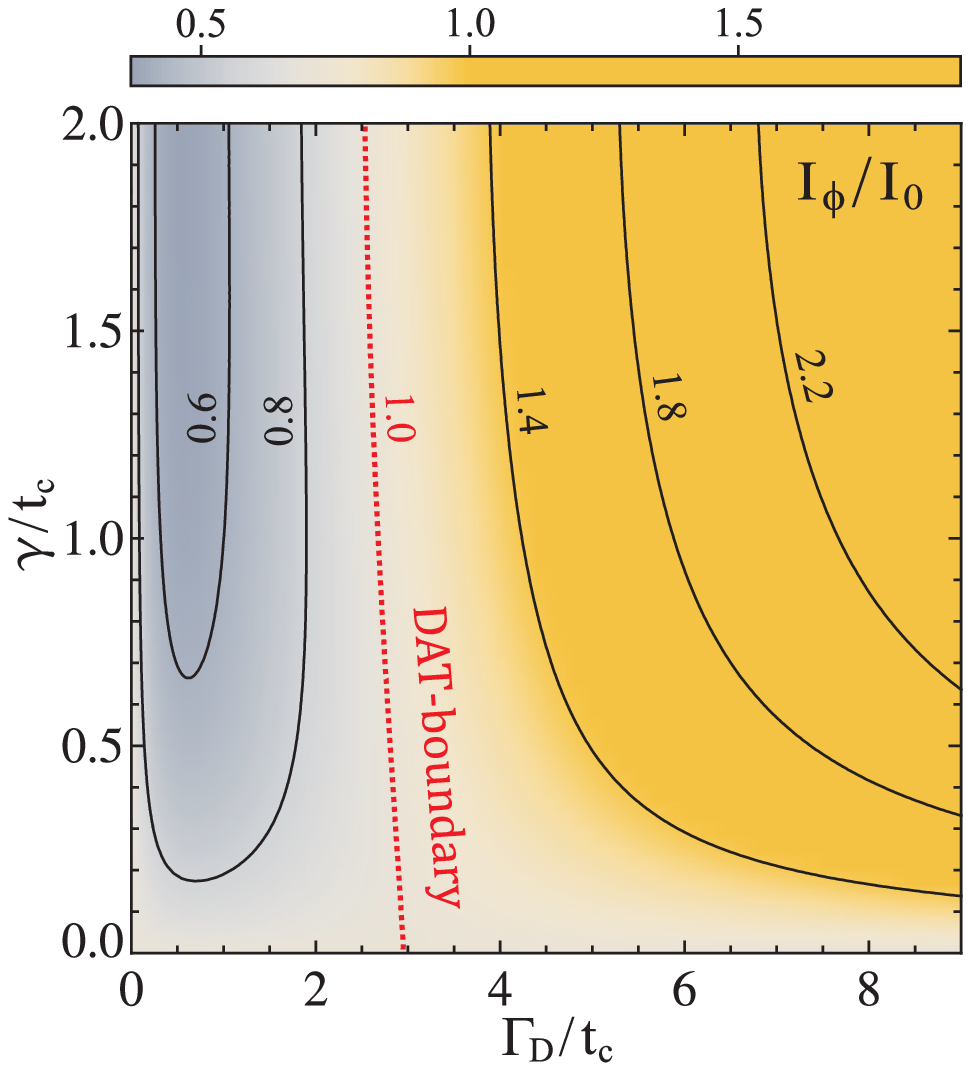}\hspace{20pt}
\raisebox{6.3cm}{{\small\bf(b)}}\includegraphics[width=170pt]{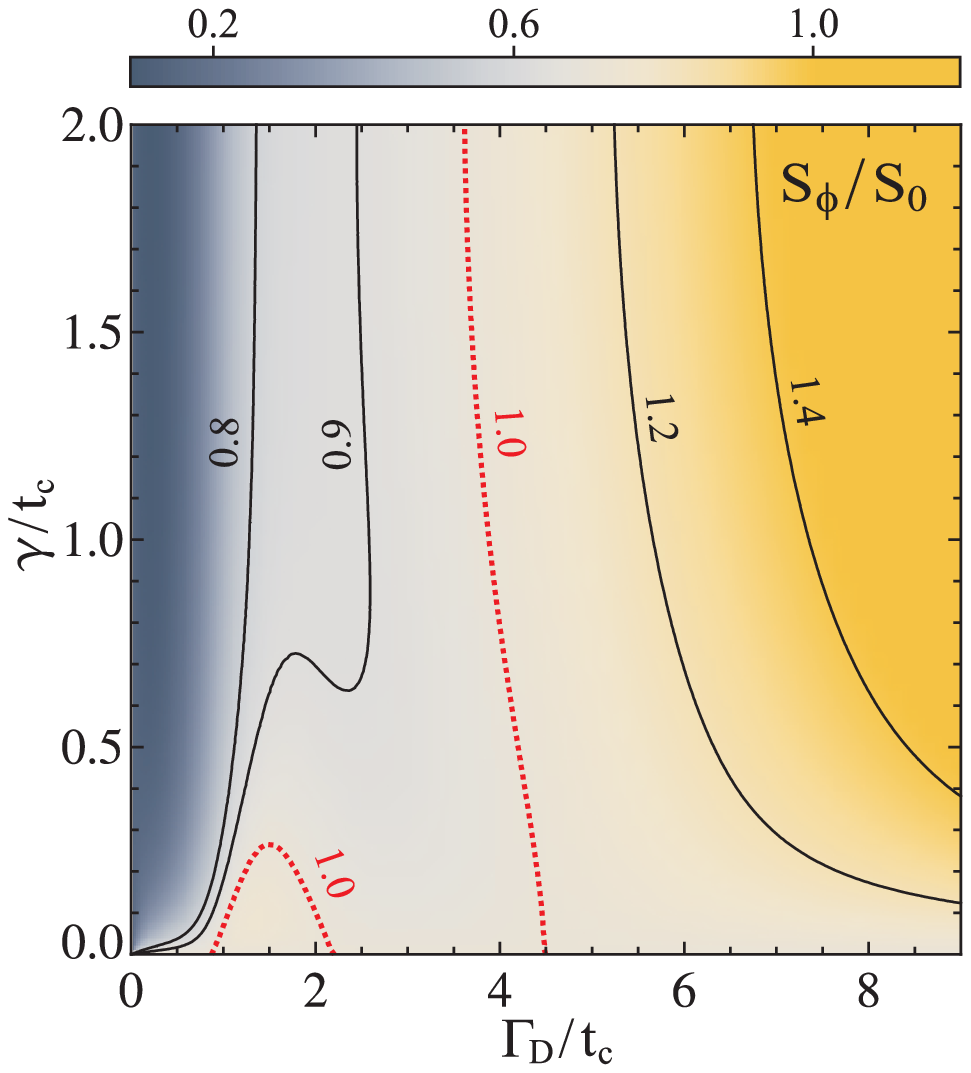}\vspace{20pt}\\
\raisebox{6.3cm}{{\small\bf(c)}}\includegraphics[width=170pt]{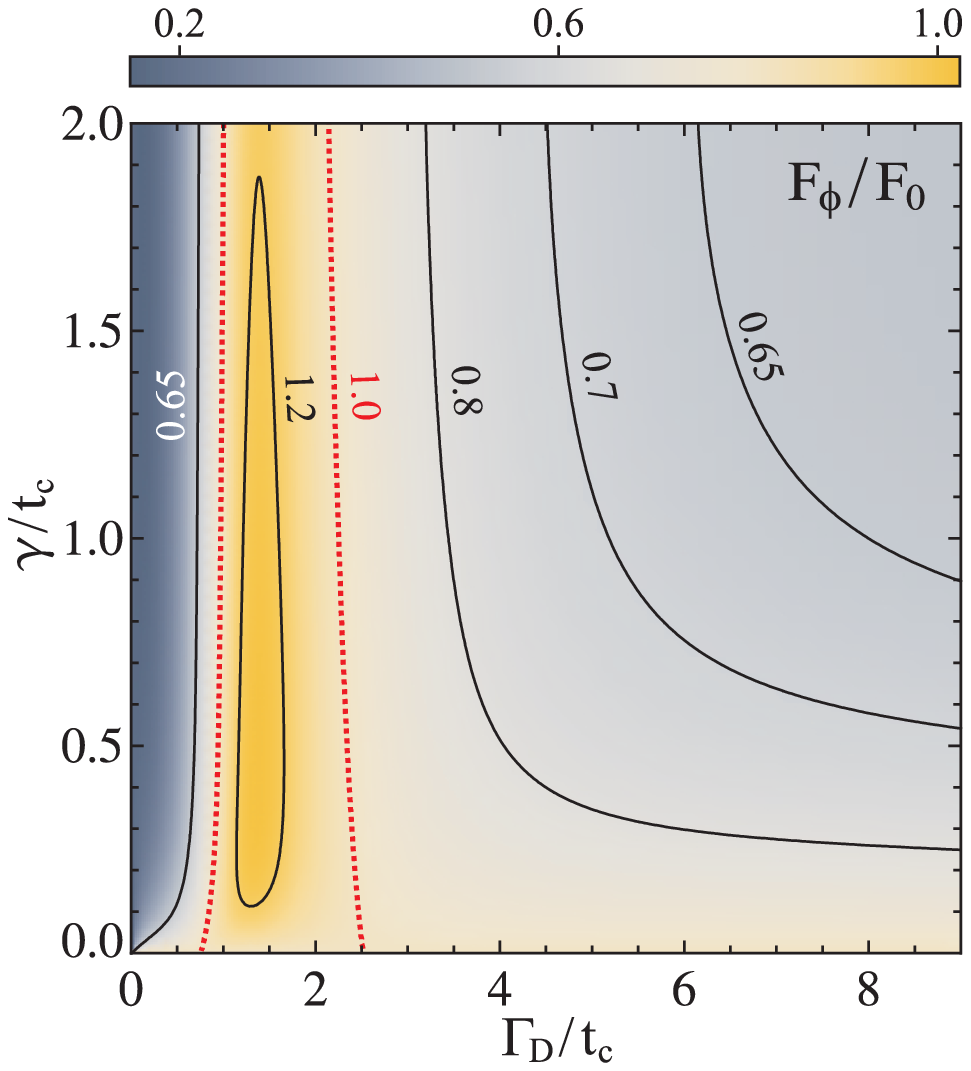}\hspace{20pt}
\raisebox{6.3cm}{{\small\bf(d)}}\includegraphics[width=171pt]{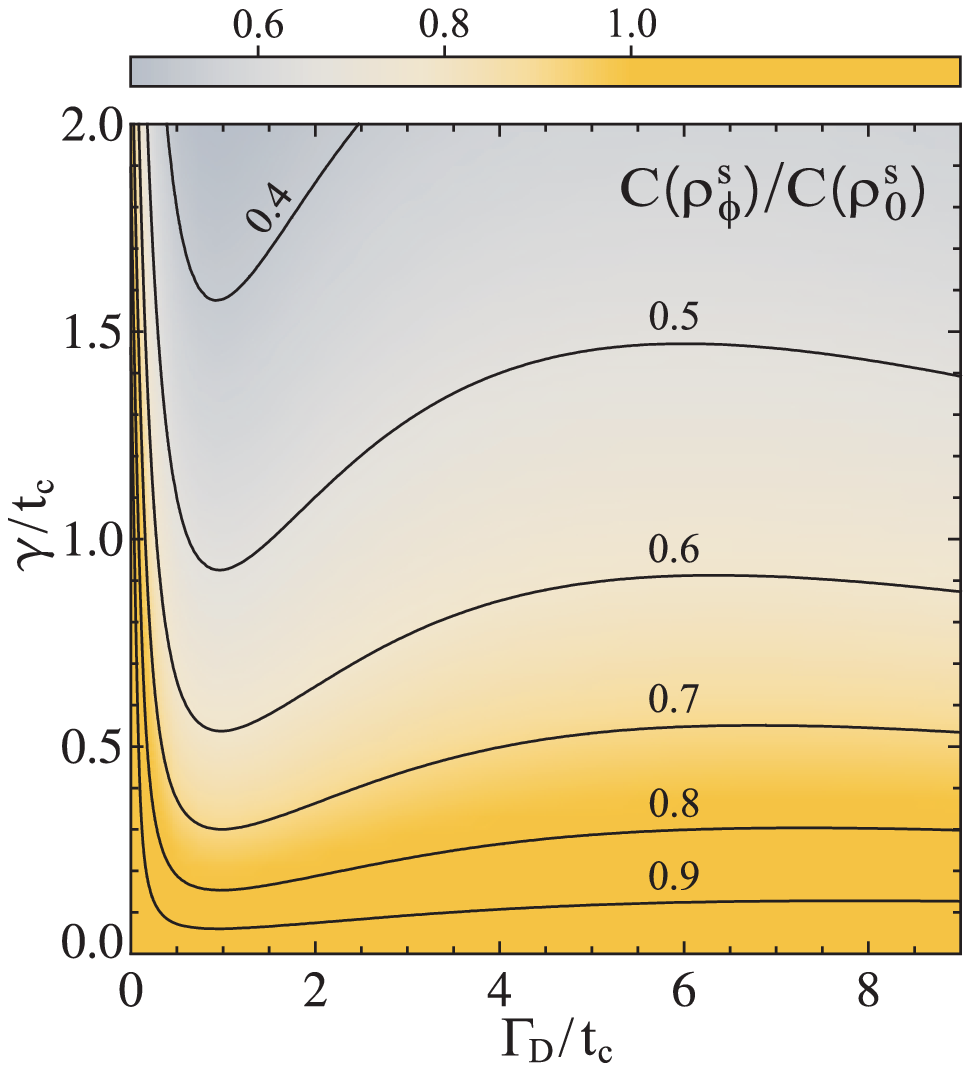}
\caption{The effect of dephasing on transport across the TQD for different dephasing rates $\gamma$ and couplings to
the drain $\Gamma_D$. {\bf(a)}~The ratio $I_\phi/I_0$ of the currents with $I_\phi$ and without $I_0$ dephasing. The
current is enhanced by dephasing ($I_\phi/I_0>1$) for $\Gamma_D$ above the threshold $\Gamma_D^*/t_c \approx 3$. The
boundary of the region with DAT ($I_\phi/I_0=1$) is given by equation~\eref{boundary}. The ratios of {\bf(b)}~the shot
noise $S_\phi/S_0$ and {\bf(c)}~the Fano factor $F_\phi/F_0$ corresponding to~{\bf(a)}. Here, the enhanced current is
accompanied by a reduced Fano factor, as seen in {\bf(c)}. {\bf(d)}~The overall coherence of the TQD is specified by
the ratio $C(\rho^s_\phi)/C(\rho^s_0)$ of the $l_1$-norm of coherence. Even moderate dephasing rates $\gamma$ reduce
the coherence considerably. There is no direct relation apparent between the overall coherence and the transport
properties. ($\Gamma_S/t_c = 1/2$ and $\var/t_c = 4$)}
\label{current}
\end{figure}

The influence of dephasing on the steady-state transport through the TQD is analyzed by
two measurable quantities, the average current $I$ and the zero-frequency shot noise $S$. More generally, we consider the
probability distribution $p(n)$ of the number $n$ of electrons that tunnel into the drain within a sufficiently
long time interval $\Delta\tau$. We use the characteristic polynomial approach~\cite{bruderer13} to determine
the full counting statistics, i.e., the distribution $p(n)$, which is parametrized
in terms of its cumulants $C_k$. The essential steps of this approach are briefly outlined in~\ref{appcpa}.
From the time-scaled cumulants $c_k=C_k/\Delta\tau$, we obtain the average
current $I=ec_1$ and the shot noise $S = 2e^2 c_2$, with $e$ the electron charge. The Fano factor $F=S/2eI$
indicates if the distribution $p(n)$ is sub-Poissonian ($F<1$) or super-Poissonian ($F>1$) and gives a measure
for the relative shot noise of the current.

\section{Dephasing-assisted transport}\label{results}

We first present the results for the transport properties of the TQD in terms of the current $I$,
shot noise $S$ and Fano factor $F$, and subsequently discuss the properties of the steady-state density
matrix $\rho^s$ of the TQD. The characteristic polynomial approach allows us to obtain the transport
properties for the most general configuration of the TQD. However, for clarity, we assume for the rest of the
paper that the tunnel couplings between the QDs are identical ($t_{12}=t_{23} = t_c$) and that QD1 and
QD3 are in resonance ($\var_1=\var_3$). The detuning $\var = |\var_1 - \var_2|$ between the central and outer
dots, identical for the $\Lambda$-type and $V$-type configuration, is then sufficient to parametrize the
on-site energies $\var_i$.

\subsection{Dephasing-assisted current and shot noise}\label{over}

The analytical result for the current across the TQD including dephasing is
\begin{equation}\label{cur}
	I_\phi = \frac{2et_c^2\Gamma_S\Gamma_D [\gamma\Gamma_\phi^2 + 2t_c^2(4\gamma + \Gamma_D)]}{D_1 + D_2D_3}\,,
\end{equation}
where the denominator is composed of
\begin{equation}\label{curparts}
\eqalign{
  D_1 &= \Gamma_S\Gamma_D\Gamma_\phi(6\gamma + \Gamma_D)\varepsilon^2\,, \\
  D_2 &= \gamma\Gamma_\phi^2 + 2t_c^2(4\gamma + \Gamma_D)\,, \\
  D_3 &= \Gamma_S\Gamma_D(3\gamma + \Gamma_D) + 2t_c^2(3\Gamma_S + \Gamma_D)\,,
}	
\end{equation}
with the rate $\Gamma_\phi = 2\gamma + \Gamma_D$. The analytical expression for the shot
noise $S_\phi$ is rather lengthy and therefore given in \ref{apptransport}. In the following
we explore the physical content of the analytical results.

To identify the parameter regime in which DAT occurs we consider the ratio between the current with
dephasing $I_\phi$ and without dephasing $I_0$, the latter being an idealization for negligible dephasing
rates. The ratio $I_\phi/I_0$ depends only weakly on $\Gamma_S$ and the detuning $\var$ is kept constant.
We therefore map out $I_\phi/I_0$ as function of $\Gamma_D$ and $\gamma$ for representative values of
$\Gamma_S$ and $\var$, with all parameters in units of $t_c$.

Figure~\ref{current}(a) shows the typical behaviour of $I_\phi/I_0$. We see that the current is significantly
enhanced by dephasing for sufficiently large values of $\Gamma_D$. More precisely, the $\gamma$-$\Gamma_D$~plane
is divided into two complementary regions, in which the current is either enhanced or suppressed by dephasing.
The boundary separating the two regions, defined by $I_\phi/I_0=1$, is determined by the relation
\begin{equation}\label{boundary}
	\var = \sqrt{6}t_c\left[\frac{\gamma(2\gamma + \Gamma_D)^2 + 2t_c^2 (4\gamma + \Gamma_D)}{\Gamma_D(2\gamma + \Gamma_D)^2 - 8t_c^2 (3\gamma + \Gamma_D)}\right]^{1/2}.
\end{equation}
We observe that equation~\eref{boundary} does not depend on $\Gamma_S$ and therefore completely defines the DAT-boundary
in the entire parameter space of the system. Moreover, for fixed detuning $\var$ and dephasing $\gamma$, we find
that equation~\eref{boundary} defines a threshold $\Gamma_D^*$ for dephasing enhanced transport, i.e., DAT is only
possible if the coupling $\Gamma_D$ exceeds the value $\Gamma_D^*$. As it can be seen in figure~\ref{current}(a),
the dependence of the threshold $\Gamma_D^*$ on the rate $\gamma$ is rather weak.

It is instructive to examine the regime of weak dephasing $\gamma\ll t_c$, which
yields particularly transparent results. Expanding $I_\phi/I_0$ to lowest order in $\gamma$ we find
\begin{equation}\label{c1expand}
	\frac{I_\phi}{I_0} = 1  + \frac{\Gamma_S\Gamma_D [\var^2(\Gamma_D^2 - 8t_c^2) - 12t_c^4]}{8t_c^6(3\Gamma_S + \Gamma_D)
	 + 2 \Gamma_S\Gamma_D^2t_c^2(2t_c^2 + \var^2)}\,\gamma\,.
\end{equation}
The prefactor of $\gamma$ can take positive or negative values. Thus, the TQD can be brought into the regime of
enhanced transport even for arbitrarily small dephasing $\gamma$ by tuning the remaining parameters. For weak
dephasing, we obtain the simple relation $\var = 2\sqrt{3}t_c^2/\sqrt{\Gamma_D^2 - 8t_c^2}$ for the DAT-boundary
and the corresponding threshold $\Gamma_D^* = 2t_c\sqrt{3t_c^2+2\var^2}/\var$, both independent of $\gamma$. This
explicit expression for $\Gamma_D^*$ shows that finite detuning $\var>0$ is necessary for DAT as $\Gamma_D^*\rightarrow\infty$
in the limit $\var\rightarrow 0$. On the other hand, in the limit of very large level detuning $\var\rightarrow\infty$
the minimum value for the threshold $\Gamma_D^*=2\sqrt{2}t_c$ is obtained.


More information about the transport properties of the TQD is encoded in the shot noise $S$.
In particular, the question arises whether or not the dephasing-assisted current comes at the
price of an increased level of shot noise. Figures~\ref{current}(b) and \ref{current}(c)
show the ratio of the shot noise $S_\phi/S_0$ and the Fano factor $F_\phi/F_0$ corresponding
to the current $I_\phi/I_0$ in figure~\ref{current}(a). We see that dephasing partly enhances
the shot noise $S_\phi/S_0$ in the DAT region; however, the level of noise relative to the
current, quantified by $F_\phi/F_0$, is reduced.

In general, the shot noise corresponding to the dephasing-assisted current depends in an intricate
way on the parameters of the system, and both dephasing-increased and -suppressed relative shot noise
may be found. To have a better understanding of the fluctuations of the current, we consider the two
specific cases of vanishingly small coupling $\Gamma_S\rightarrow 0$ and large coupling $\Gamma_S\gg t_c$
to the source. In the former limit, we expand $F_\phi/F_0$ to lowest order in $\Gamma_S$ to find

\begin{equation}\label{fgamma}
	\frac{F_\phi}{F_0} = 1 - \left[\frac{3\gamma}{t_c^2} + \frac{\gamma\var^2(24 \gamma t_c^2 + 8\Gamma_D t_c^2
	- \Gamma_D\Gamma_\phi^2)}{2t_c^4(8 \gamma t_c^2 + 2\Gamma_D t_c^2+\gamma\Gamma_\phi^2)}\right]\Gamma_S\,.
\end{equation}
From equation~\eref{fgamma} we determine the boundary between enhanced and suppressed relative shot noise, defined
by $F_\phi/F_0=1$. It turns out that this boundary is identical to the DAT-boundary given by equation~\eref{boundary}
and that enhanced transport is always accompanied by an increased Fano factor in the limit
$\Gamma_S\rightarrow 0$.

In the opposite limit $\Gamma_S\gg t_c$, we expand $F_\phi/F_0$ to lowest order in $\Gamma_S^{-1}$ to obtain
an expression for the boundary $F_\phi/F_0=1$ which is independent of $\Gamma_S$. In contrast to the
previous case, the DAT region lies inside the region of a reduced Fano factor for all dephasing rates $\gamma$,
similar to the example shown in figures~\ref{current}(a) and \ref{current}(c). Thus, dephasing not only enhances
the current through the TQD but also reduces the relative fluctuations in the regime $\Gamma_S\gg t_c$.

\subsection{Occupations and coherences of the reduced density matrix for the TQD}\label{coh}

We now turn to the discussion of the steady-state density matrix $\rho^s$ of the TQD, which is determined
by solving $\gen\rho^s = 0$. The occupations $\rho^s_{ii}$ and especially the coherences between different
electronic states $\rho^s_{i\neq j}$ play a crucial role for the transport processes. Unlike the current
and shot noise, the density matrix $\rho^s$ is not directly accessible by measurements, yet useful for
complementing our previous findings in terms of the dynamics of the TQD.

The occupation probabilities for each state of the TQD including dephasing read
\begin{equation}\fl\label{eqs_pop}
\eqalign{
\rho_{11}^s &= \frac{\Gamma_S}{2}\,\frac{2t_c^2(8\gamma^3 + 24\gamma^2\Gamma_D + 10\gamma\Gamma_D^2 + \Gamma_D^3) + (4\gamma+\Gamma _D)[8t_c^4
	+\Gamma_{\phi}\Gamma_D(\gamma\Gamma_{\phi}+2\epsilon^2)]}{D_1+D_2D_3}\,, \\
\rho_{22}^s &= \frac{\Gamma_S}{2}\,\frac{2t_c^2\Gamma_{\phi}(4\gamma^2+6\gamma\Gamma_D+\Gamma_D^2)
		+ 8 t_c^4(4\gamma +\Gamma_D)+\gamma\Gamma_{\phi}\Gamma_D(\Gamma_{\phi}^2+4\epsilon ^2)}{D_1+D_2D_3}\,, \\
\rho_{33}^s &=\frac{\Gamma_S}{2}\,\frac{4t_c^2[\gamma\Gamma_{\phi}^2+2t_c^2(4\gamma +\Gamma_D)]}{D_1+D_2D_3}\,,
}
\end{equation}
where $D_1$, $D_2$ and $D_3$ are defined in equation~\eref{curparts}. The occupation of the empty state $\rho_{00}^s$,
which is typically small but finite, can be found from the normalization condition $\sum_{i} \rho_{ii}^s=1$.
Note that in our case of infinite bias voltage and strong Coulomb blockade, the steady-state current measured at
the drain is given by $I=e\Gamma_D\rho_{33}^s$.

\begin{figure}[t]
\centering
\raisebox{5.7cm}{{\small\bf(a)}}\includegraphics[width=171pt]{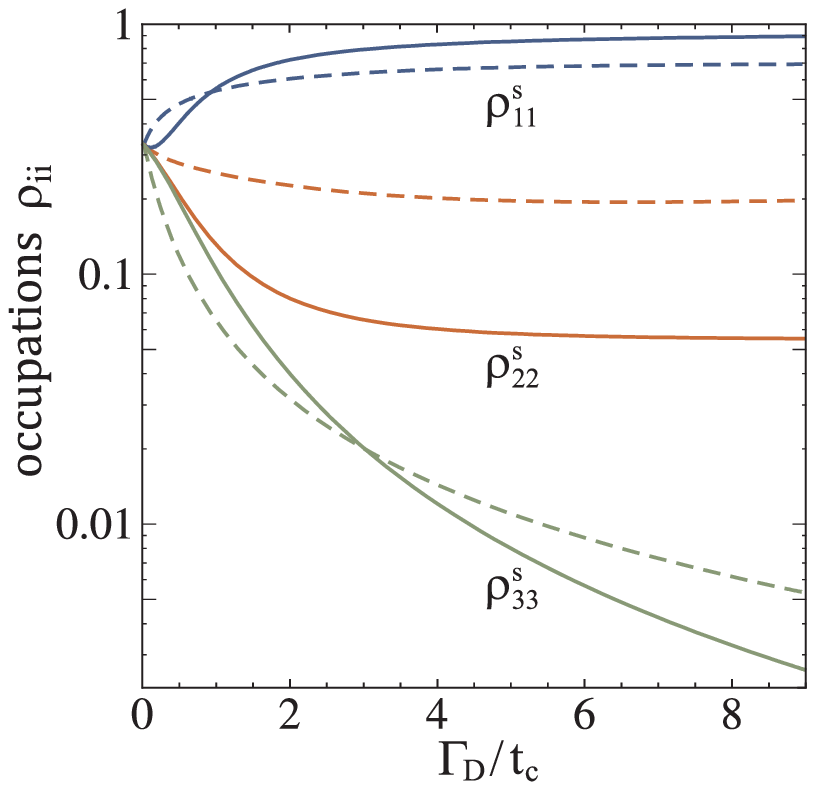}\hspace{30pt}
\raisebox{5.7cm}{{\small\bf(b)}}\includegraphics[width=170pt]{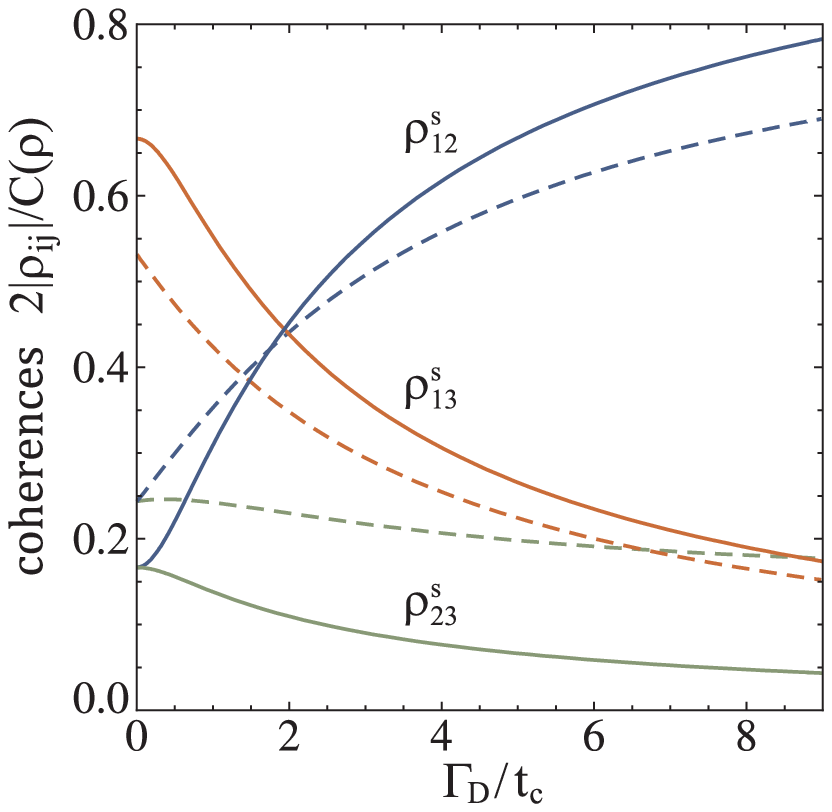}
\caption{The effect of dephasing on the properties of the steady-state density matrix $\rho^s$ of the TQD. {\bf(a)}~The occupations $\rho_{ii}^s$
as a function of the coupling to the drain $\Gamma_D$ without (solid line) and with (dashed line) dephasing. Dephasing increases $\rho_{33}^s$ for
$\Gamma_D>\Gamma_D^*$ and reduces the difference between $\rho_{11}^s$ and $\rho_{22}^s$. {\bf(b)}~The weights of the coherences $\rho_{ij}^s$ with
respect to the overall coherence $C(\rho^s)$. The large weight of $\rho_{13}^s$ for $\Gamma_D\lesssim t_c$ indicates long-distance tunnelling and
the dominance of $\rho_{12}^s$ for $\Gamma_D\gg t_c$ is a signature for the quantum Zeno effect. 
($\Gamma_S/t_c = 1/2$, $\var/t_c = 4$ and $\gamma/t_c=2/3$)}
\label{coherence}
\end{figure}

The occupations $\rho_{ii}^s$ for increasing $\Gamma_D$ with and without dephasing are shown in figure~\ref{coherence}(a).
In accord with the behaviour of the current, dephasing reduces the occupation $\rho_{33}^s$ of QD3 in the regime
$\Gamma_D < \Gamma_D^*$, while $\rho_{33}^s$ is dephasing-enhanced in the regime $\Gamma_D > \Gamma_D^*$ where DAT is
observed. In the fully coherent case, i.e., in absence of dephasing, both $\rho_{11}^s$ and $\rho_{22}^s$ show signatures
of the trapped coherent evolution between QD1 and QD2: For sufficiently large values of $\Gamma_D$, the occupations
$\rho_{11}^s$ and $\rho_{22}^s$ are approximately constant and account for most of the occupation of the TQD. In addition,
we observe that $\rho_{22}^s\ll\rho_{11}^s$ as a result of the large detuning $\var$ between the QDs, such that the charge
is essentially localized in QD1. Dephasing significantly reduces the difference between all occupations $\rho_{ii}^s$, in
particular between $\rho_{22}^s$ and $\rho_{11}^s$, and consequently counteracts the localization effect. We will further
clarify the behavior of the occupations in context of the quantum Zeno effect in the next section.

The coherences of the TQD are given by
\begin{equation}\label{eqs_coh}
\eqalign{
\rho_{12}^{\rm s} &= \frac{t_c \Gamma _D\Gamma _S [(\var + \rmi\gamma)\Gamma_{\phi}^2 + 2\rmi t_c^2(4 \gamma + \Gamma _D)]}{D_1 + D_2 D_3}\,, \\
\rho_{23}^{\rm s} &= \frac{\rmi t_c\Gamma_D\Gamma_S [2t_c^2(4 \gamma +\Gamma_D) + \gamma\Gamma_{\phi}(\Gamma_{\phi} + 2\rmi\var)]}{D_1 + D_2 D_3}\,, \\
\rho_{13}^{\rm s} &= \frac{2\rmi\var t_c^2 \Gamma_D \Gamma_S (4\gamma + \Gamma_D)}{D_1 + D_2 D_3}
}
\end{equation}
and $\rho_{j0}^{\rm s} = 0$ for $j\neq 0$. To assess the
effect of dephasing on the coherence of the TQD in a quantitative way we use the $l_1$-norm~\cite{baumgratz13}
\begin{equation}\label{norm}
		C(\rho) = \sum_{i\neq j} |\rho_{ij}|
\end{equation}
as a measure of the overall coherence in the system. Specifically, we analyze the ratio $C(\rho^s_\phi)/C(\rho^s_0)$,
shown in figure~\ref{current}(d) as a function of $\Gamma_D$ and $\gamma$. As expected, we see that the overall coherence
of the TQD is reduced with increasing $\gamma$. It is interesting, however, that for constant $\gamma$ the coherence tends to
be higher in the DAT region, which indicates that dephasing-enhanced transport is not directly related to a reduction of
the overall coherence.

The weights of specific coherences $\rho^s_{ij}$ with respect to the overall coherence $C(\rho)$ are more informative and can be quantified
as $2|\rho^s_{ij}|/C(\rho^s)$. This ratio is shown in figure~\ref{coherence}(b) for the coherences $\rho^s_{13}$, $\rho^s_{12}$ and $\rho^s_{23}$
with and without dephasing as a function of $\Gamma_D$. We see that the coherence $\rho^s_{13}$, indicating
long-distance tunnelling between the outer QDs of the array, accounts for most of the coherence in the regime
$\Gamma_D\lesssim t_c$ while $\rho^s_{12}$ is the dominating contribution for $\Gamma_D\gg t_c$. These observations
are discussed further in the next section and support our explanations of DAT and suppressed transport based on the
quantum Zeno effect and long-distance tunnelling, respectively.

\section{Underlying dynamic processes}\label{under}

We now focus on the underlying physical processes that account for enhanced and suppressed currents
across the system. Transport through the TQD is determined by its energy levels, the tunnel coupling
and the effects of the reservoirs. Accordingly, DAT can be explained as a consequence of level-broadening
caused by both, pure dephasing and the coupling to the reservoirs. Nevertheless, our previous results
suggest that we can also understand DAT in terms of the dynamics of the charge, which yields further
insight into the role of dephasing. To this end, we present two effective models for the evolution of
the TQD, which involve the equivalent of the quantum Zeno effect in quantum transport~\cite{gurvitz98,chen04}
and long-distance tunnelling~\cite{busl13,vandersypen13,sanchez14b}. Both models reduce the problem
of the TQD to an effective double quantum dot (DQD).

\subsection{Quantum Zeno effect}\label{zenoo}

Dephasing-enhanced transport occurs for $\Gamma_D$ exceeding the threshold $\Gamma_D^*$, typically $\Gamma_D \gg t_c$, provided
that the detuning $\var$ is finite. To understand the predicted DAT let us first consider the TQD in the fully coherent regime.
The coherent evolution of the electrons in the TQD is characterized by the time scale $\tau_c\sim t_c^{-1}$. Moreover, because
QD3 is coupled to the drain we have that QD3 is repeatedly emptied, or equivalently observed, on a time scale $\tau_D\sim\Gamma_D^{-1}$,
the average time interval between subsequent detections. For sufficiently strong coupling $\Gamma_D$, the irreversible tunnelling
into the drain can be considered as a continuous measurement of QD3 on a time scale $\tau_D\ll\tau_c$, so that the ensuing
Zeno effect restricts the evolution of the charge to QD1 and QD2.\footnote{Note that within our assumptions for the reservoirs
(wide-band limit and infinite bias voltage) our model is not restricted to small values of $\Gamma_D$~\cite{prachavr10}.}
In other words, the charge is localized and evolves coherently between QD1 and QD2 only, which is confirmed by the dominance of
the coherence $\rho_{12}^s$ in the regime $\Gamma_D\gg t_c$ [cf.~figure~\ref{coherence}(b)].

To be more quantitative, we describe the TQD in the Zeno regime in absence of dephasing by an effective model
for QD1 and QD2 whose evolution is governed by
\begin{equation}\label{zeno}
\eqalign{
  H^{\rm Z} &=  (t_{12}\cre{d}_2\an{d}_1 + \hc) + \sum_{i=1,2}\var^{\rm Z}_i\cre{d}_i\an{d}_{i}\,, \\
	\gen_D^{\rm Z}\rho &= \frac{\Gamma_D^{\rm Z}}{2}(2\an{d}_2\rho\cre{d}_2 - \{\cre{d}_2\an{d}_2,\rho\})\,,
}	
\end{equation}
together with the Lindblad operator $\gen_S$ in equation~\eref{super}. The evolution of the charge in
this effective DQD is reduced to the subspace spanned by the states $\{\ket{0},\ket{1},\ket{2}\}$
and tunnelling into the drain occurs now from QD2 with the effective rate~\cite{facchi03,pascazio09}
\begin{equation}\label{effcoup}
	\Gamma_D^{\rm Z} = 2t_{23}^2\frac{\Gamma_D/2}{\left(\Gamma_D/2\right)^2 + (\var_2-\var_3)^2}\,.
\end{equation}
In addition, the on-site energy of the second QD is shifted to $\var^{\rm Z}_2 = \var_2(1+ \Gamma_D^{\rm Z}/\Gamma_D)$,
with however a negligible effect on the transport properties, while $\varepsilon_1$ is unaffected.
We note that, alternatively, equations~\eref{zeno} and \eref{effcoup} can be interpreted as a strong Lorentzian level broadening
of QD3 due to the coupling to the drain with the corresponding reduction of the local spectral density. The effective Zeno model
is valid in the regime $\Gamma_D\gg t_c$ and $\gamma=0$, where it reproduces the behaviour of the current through the TQD, as shown
in \ref{appzeno}.

Importantly, the current is strongly suppressed as the dynamics approaches the Zeno regime, according to equation~\eref{effcoup},
because the effective coupling to the drain $\Gamma_D^{\rm Z}\ll\Gamma_D$ constitutes a transport bottleneck. We also
point out the essential fact that the Zeno effect relies on the coherent evolution of the TQD. Therefore, the transport
bottleneck is not expected to occur for incoherent transport between the QDs.

The role of dephasing is to impair the coherence of the TQD and hence to render the Zeno effect
less efficient. In fact, dephasing reduces the trapped coherent evolution between QD1 and QD2,
resulting in the reduction of the coherence $\rho^s_{12}$ [cf.~figure~\ref{coherence}(b)], and generally
enables incoherent transitions between all QDs. In this way, dephasing partially alleviates the transport
bottleneck and consequently the dephasing-assisted current arises. This is reflected, also, in the
change of the occupations $\rho^s_{ii}$ due to the presence of dephasing, as specifically $\rho^s_{11}$ is
reduced while $\rho^s_{22}$ and $\rho^s_{33}$ are increased for finite $\gamma$ [cf.~figure~\ref{coherence}(a)].

\subsection{Long-distance tunnelling}\label{long}

The suppression of the current for $\Gamma_D\lesssim t_c$, as seen in figure~\ref{current}(a), can be understood in
terms of the long-distance tunnelling between QD1 and QD3. For the large detuning $\var$ of
the central QD, long-distance tunnelling provides the dominant mechanism for charge transport through the TQD.
The effect of dephasing is to destruct this coherent pathway, leading to a suppressed current.

In order to have a better understanding of the negative impact of dephasing, we use an effective model for QD1 and QD3.
In the coherent regime and for large detuning $\var\gg t_c$ we can eliminate the state $\ket{2}$ from the dynamics
of the TQD, leaving us with an effective DQD described by
\begin{equation}\label{lrange}
	H^{\rm ld} = (t_{13}^{\rm ld}\cre{d}_3\an{d}_1 + \hc) + \sum_{i=1,3}\var^{\rm ld}_i\cre{d}_i\an{d}_{i}\,,
\end{equation}
together with the Lindblad operators $\gen_S$, $\gen_D$ and $\gen_\phi$ in equation~\eref{super}. Here, $t_{13}^{\rm ld}$
is the effective long-distance tunnelling between QD1 and QD3, $\var^{\rm ld}_i$ are renormalized on-site
energies and the parameters entering $\gen_S$, $\gen_D$ and $\gen_\phi$ are replaced by their effective counterparts
$\Gamma_S^{\rm ld}$, $\Gamma_D^{\rm ld}$ and $\gamma^{\rm ld}$. Note that the evolution of the charge is reduced now
to the subspace spanned by the states $\{\ket{0},\ket{1},\ket{3}\}$. The effective long-distance model
reproduces the current across the TQD for $\gamma=0$ and $\Gamma_D\lesssim t_c$, as shown in \ref{appld},
where we used $t_{13}^{\rm ld}=(\varepsilon-\sqrt{\varepsilon^2+8t_c^2})/4$ obtained from an improved adiabatic
elimination method~\cite{paulisch14}.

In general, for this effective DQD we find that dephasing always leads to a suppression of the current provided that
$\var^{\rm ld}_1=\var^{\rm ld}_3$, as in this case [cf.~\ref{appld}]
\begin{equation}\label{suppress}
	 \frac{I_\phi^{\rm ld}}{I_0^{\rm ld}} = 1 - \frac{2\gamma^{\rm ld}\Gamma_S^{\rm ld}\Gamma_D^{\rm ld}}{\Gamma_S^{\rm ld}\Gamma_D^{\rm ld}(2 \gamma^{\rm ld}
			+ \Gamma_D^{\rm ld}) + (2t_{13}^{\rm ld})^2(2\Gamma_S^{\rm ld} + \Gamma_D^{\rm ld})}\,,
\end{equation}
so that indeed $I_\phi^{\rm ld}/I_0^{\rm ld}\leq 1$. This result, which is independent of the details of the effective model,
explains the suppressed current of the TQD in regimes where long-distance tunnelling plays an important role in transport.
The reduction of the current can be significant even for weak dephasing $\gamma\ll t_c$ as the relevant ratio
$\gamma^{\rm ld}/t_{13}^{\rm ld}$ may still be large.

\section{Experimental implementation}\label{exp}
\begin{figure}[t]
\centering
\raisebox{5.65cm}{{\small\bf(a)}}\includegraphics[width=170pt]{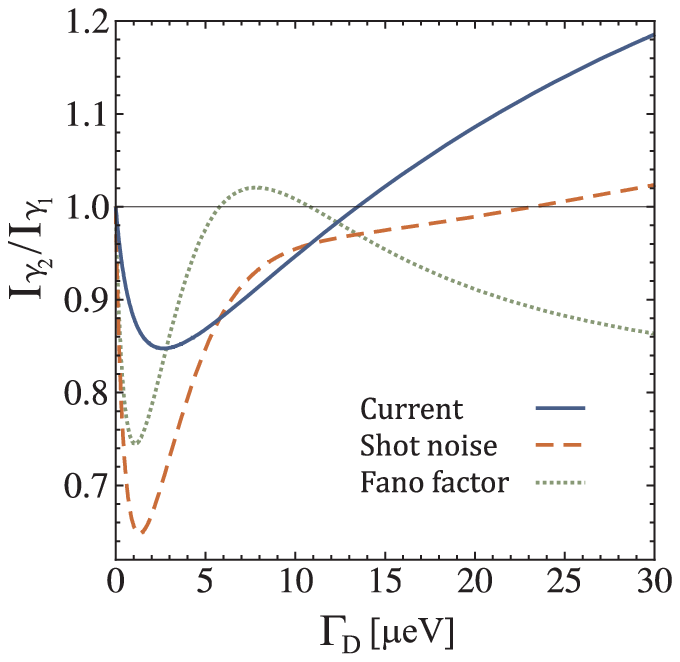}\hspace{30pt}
\raisebox{5.65cm}{{\small\bf(b)}}\includegraphics[width=167pt]{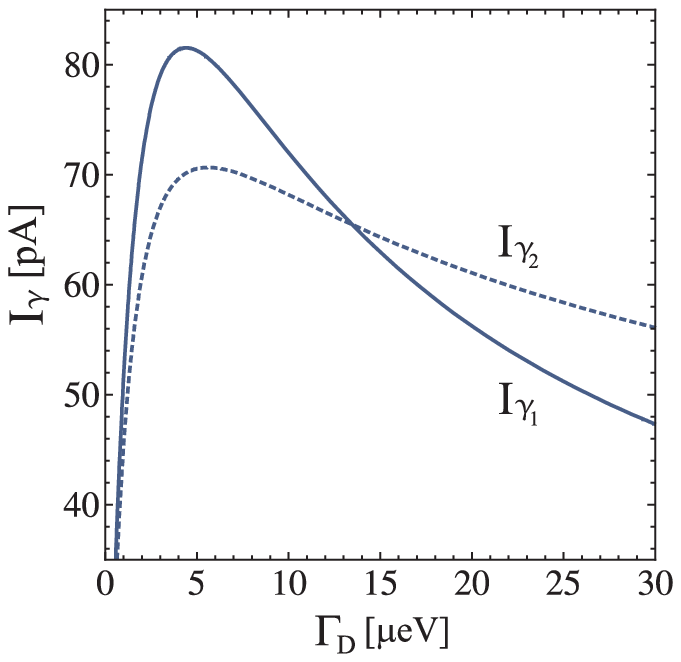}
\caption{{\bf(a)}~The ratio of the currents $I_{\gamma_2}/I_{\gamma_1}$ (solid line) for two values of the dephasing rate $\gamma_2>\gamma_1$.
For $\Gamma_D$ exceeding the threshold $\Gamma_D^*\cong 13.5\ev$ we observe dephasing-enhanced transport, i.e., $I_{\gamma_2}/I_{\gamma_1}>1$.
The shot noise $S_{\gamma_2}/S_{\gamma_1}$ (dashed line) increases while the Fano factor $F_{\gamma_2}/F_{\gamma_1}$ (dotted line) decreases
in the DAT region. {\bf(b)}~The currents $I_{\gamma_1}$ (solid line) and $I_{\gamma_2}$ (dashed line) for
increasing $\Gamma_D$. The crossing at the threshold $\Gamma_D^*$ provides a clear signature for the detection of DAT.
(Parameters in the main text)}
\label{explot}
\end{figure}

Let us now discuss how DAT can be unambiguously detected in an experiment. An interesting aspect of transport through
the linear TQD is not only that it exhibits DAT but, importantly, that the boundary between suppressed and enhanced
transport can be crossed by changing a single external parameter. This crossing provides a clear signature for the
observation of DAT.

In order to detect DAT we envisage the following experimental procedure with all parameters based on typical
values for lateral QDs~\cite{hayashi03,busl13,vandersypen13}: The current and shot noise are measured
with two different values of the dephasing rate, $\gamma_1=1\ev$ and $\gamma_2=2.5\ev$. In both measurements,
the coupling of QD3 to the drain $\Gamma_D$ is varied in the range $0-30\ev$, while the inter-dot tunnelling coupling
$t_c=5\ev$, the detuning $\var=20\ev$ and the coupling to the source $\Gamma_S=1\ev$ are kept constant.

The dephasing rate $\gamma$ can be modified, for instance, by increasing the temperature of the environment,
which results in enhanced electron-phonon interactions~\cite{fedichkin05}. For example, in the case of
piezoelectric phonons~\cite{brandes99,brandes02} the required values of the dephasing rates $\gamma_1$ and
$\gamma_2$ can be achieved for phonon temperatures of approximately $1{\rm K}$ and $3{\rm K}$, respectively.
We emphasize that the exact value of the rates $\gamma_1$ and $\gamma_2$ is not essential for our scheme as
long as $\gamma_2>\gamma_1$, and that the strength of dephasing has not to be identical for all QDs, i.e.,
unavoidable small fluctuations of the dephasing rate $\delta\gamma\ll\gamma_2-\gamma_1$ acting on each QD do
not affect the overall qualitative results.

Considering the ratio of the measured currents $I_{\gamma_2}/I_{\gamma_1}$ as a function of the coupling
to the drain $\Gamma_D$, we expect to observe the crossover to DAT close to $\Gamma_D\cong 13.5\ev$, as shown
in figure~\ref{explot}(a). The shot noise for $\gamma_1$ and $\gamma_2$ is practically identical in the DAT
regime so that the ratio of the Fano factors $F_{\gamma_2}/F_{\gamma_1}$ decreases for the selected parameters.
The specific values of the two currents $I_{\gamma_1}$ and $I_{\gamma_2}$, which are more relevant for the
experimental implementation, are shown in figure~\ref{explot}(b). The currents deviate by approximately $10$pA ($15\%$)
well inside the regions where transport is either suppressed or enhanced, which is detectable by using standard
technologies for lateral quantum dots.

\section{Conclusions}\label{conc}

We have proposed a concrete realization of DAT in a highly controllable quantum system. Our comprehensive
and fully analytical results demonstrate that dephasing-assisted currents can be observed
in linear TQDs in a transport configuration. Moreover, we have suggested an experimental procedure that provides
an unambiguous way to evidence the occurrence of DAT under controlled conditions.

In particular, we have found that enhanced transport in linear TQDs occurs for sufficiently strong coupling to the
drain and results from the reduction of the quantum Zeno effect, in contrast to triangular TQDs, where destructive
interferences are the underlying cause for DAT. We also found that dephasing can suppress the current for moderate
coupling to the drain due to the reduction of the long-distance tunnelling between the outer dots of the TQD. Both
enhancement and suppression of the current induced by dephasing were shown to be directly reflected in the behavior
of the coherences of the reduced density matrix of the TQD. While the above results have been obtained for a symmetric
configuration of the linear TQD our findings are qualitatively valid also for moderately asymmetric configurations, e.g.,
with slightly different tunnelling couplings, $t_{12}\neq t_{23}$. In such a case, significant changes of the transport
properties occur only in the regime of vanishingly weak coupling to the drain.

The example of the TQD clearly shows that the coupling to the drain reservoir plays a crucial role in DAT. This general
conclusion may be relevant for other transport systems, in particular for larger arrays of QDs. It is worth mentioning
that our results could be used to extract the dephasing rate $\gamma$ from the transport properties of a linear array of QDs,
in particular from the electrical current. Finally, we have revealed that dephasing induces significant changes
in the relative level of shot noise. The underlying mechanisms responsible for these changes and the consequences of this
result need yet to be explored.


\ack
We are grateful to R. Aguado and G. Platero for helpful discussions and comments.
This work has been supported by the ERC Synergy grant BioQ, the EU
Integrating project SIQS, the EU STREP project PAPETS and the Alexander
von~Humboldt Foundation.


\appendix

\section{Characteristic polynomial approach}\label{appcpa}

The characteristic polynomial approach to determine the full counting statistics for an open quantum system is explained in
detail in reference~\cite{bruderer13}. Let us briefly outline here the essential steps of the method.
We first transform the original $\gen$ in equation~\eref{blad} into the deformed Lindblad operator $\gen\rightarrow\gen_\xi$ by
making the substitution $\an{d}_3\rho\cre{d}_3\rightarrow\rme^{\xi}\an{d}_3\rho\cre{d}_3$ in equation~\eref{super},
thereby introducing the counting variable $\xi$. We then calculate the characteristic polynomial
$P_\xi(x) = \det[x\mathcal{I}-\gen_\xi]$ of the deformed operator $\gen_\xi$, where $\gen_\xi$ is expressed
as a matrix of dimension $M\times M$ and $\mathcal{I}$ is the identity. The characteristic polynomial $P_\xi(x)$
in its coefficient form reads
\begin{equation}\label{poly}
	P_\xi(x) = x^M + a_{M-1}x^{M-1} + \sum_{j=0}^{M-2}a_j(\xi)x^j\,.
\end{equation}
The coefficients $a_j = a_j(\xi)|_{\xi=0}$ and their derivatives $a_j^{\prime} = \partial_\xi a_j(\xi)|_{\xi=0}$,
$a_j^{\prime\prime} = \partial_\xi^2 a_j(\xi)|_{\xi=0}$ and so forth
immediately yield the time-scaled cumulants $c_k=C_k/\Delta\tau$ of the distribution $p(n)$. Explicitly,
we have for the average $c_1$, the variance $c_2$ and the Fano factor $F\equiv c_2/c_1$
\begin{equation}\label{ics}
\eqalign{
c_1 &= -\frac{a_0^{\prime}}{a_1}\,, \\
c_2 &= -\frac{1}{a_1}(a_0^{\prime\prime} + 2a_1^{\prime}c_1 + 2a_2c_1^2)\,, \\
F &= \frac{a_0^{\prime\prime}}{a_0^{\prime}} + \frac{2}{a_1^2}(a_0^{\prime}a_2 - a_1^{\prime} a_1)\,. 
}
\end{equation}
The expressions above can be further simplified as $a_j^{\prime\prime}=a_j^{\prime}$ for our specific set-up with
a single source and drain in the infinite bias limit. From the cumulants we directly obtain analytical expressions
for the average current $I=ec_1$ and the zero-frequency shot noise $S = 2e^2 c_2$, with $e$ the electron charge.

\section{Transport properties of the TQD including pure dephasing}\label{apptransport}

\begin{figure}[b]
\centering
\raisebox{5.2cm}{{\small\bf(a)}}\includegraphics[width=160pt]{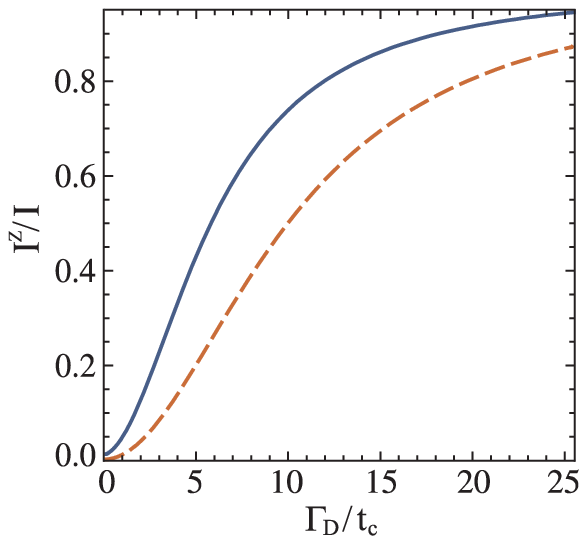}\hspace{35pt}
\raisebox{5.2cm}{{\small\bf(b)}}\includegraphics[width=160pt]{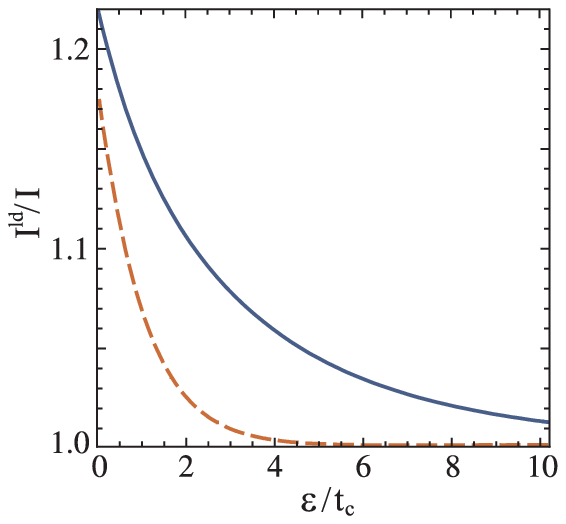}
\caption{{\bf (a)}~The ratio $I^{\rm Z}/I$ for two values of the level detuning, $\var/t_c=3$ (solid line) and
$\var/t_c=5$ (dashed line), showing that the agreement between the effective DQD model, equation~\eref{zeno},
and the full TQD system is reached for $\Gamma_D\gg t_c$.
{\bf (b)}~The ratio $I^{\rm ld}/I$ for two values of the coupling to the drain reservoir,
$\Gamma_D/t_c=1$ (solid line) and $\Gamma_D/t_c=3/2$ (dashed line), showing the agreement
between the effective DQD model for QD1 and QD3, equation~\eref{lrange}, and the full TQD
system in the regime $\var\gg t_c$. ($\Gamma_S/t_c=1/2$ and $\gamma=0$)}
\label{zenoplot}
\end{figure}

To obtain the general expressions for $c_1$ and $c_2$ we first construct the matrix of dimension $16\times 16$
corresponding to $\gen_\xi$, determine the characteristic polynomial $P_\xi(x)$ and then make use of equations~\eref{ics}.
The result for the current $c_1$ is given in the main text [cf.~equation~\eref{cur}]. For the shot noise we find
\begin{equation}\label{c2}
	c_2 = \frac{B_1}{B_9}\bigg(B_2B_3 + \sum_{i=4}^8 B_i\bigg),
\end{equation}
where the expressions for the $B_i$ are
\begin{equation}\label{c2parts}
\eqalign{
  B_1 &= 2\Gamma_S \Gamma_D t_c^2 [\gamma\Gamma_\phi^2 + 2 t_c^2 (4 \gamma +\Gamma_D)]\,, \\
	B_2 &= 2\Gamma_S^2 \Gamma_D t_c^2\Gamma_\phi^2\,, \\
	B_4 &= 16t_c^8 (4 \gamma +\Gamma_D)^2 [2 \gamma(9 \Gamma_S^2+\Gamma_D^2)
	+11 \Gamma_S^2 \Gamma_D+\Gamma_D^3]\,,
}	
\end{equation}
with the rate $\Gamma_\phi = 2\gamma + \Gamma_D$, and furthermore
\begin{equation}\label{c2longparts}
\hspace{-20pt}\eqalign{
  B_3 &= \gamma\Gamma_\phi(40 \gamma^4+92\gamma^3 \Gamma_D +60 \gamma^2 \Gamma_D^2 +17 \gamma \Gamma_D^3+2 \Gamma_D^4) +8 \gamma \Gamma_D \var^4\\
	&\quad+\var^2 (80 \gamma^4+264 \gamma^3 \Gamma_D+164 \gamma^2 \Gamma_D^2+34 \gamma  \Gamma_D^3+3 \Gamma_D^4)\,, \\
	B_5 &= \Gamma_S^2 \Gamma_D^2\Gamma_\phi^3[\var^4 (28 \gamma^2+8\gamma \Gamma_D+\Gamma_D^2)+\gamma\var^2\Gamma_\phi(28 \gamma^2+14\gamma\Gamma_D+\Gamma_D^2)\\
	&\quad+\gamma^2\Gamma_\phi^2 (7 \gamma^2+5 \gamma  \Gamma_D+\Gamma_D^2)] \,, \\
	B_6 &= 8t_c^6 \Gamma_\phi(4 \gamma +\Gamma_D)[8 \gamma^3(9 \Gamma_S^2+\Gamma_D^2 )+8 \gamma^2  (15 \Gamma_S^2 \Gamma_D+\Gamma_D^3)\\
	&\quad+2\gamma(14 \Gamma_S^2 \Gamma_D^2+\Gamma_D^4 )-\Gamma_S^2 \Gamma_D  (\Gamma_D^2-4 \var^2 )] \,, \\
	B_7 &= 4t_c^4\Gamma_\phi[16\gamma^6  (9 \Gamma_S^2+\Gamma_D^2 ) + 16\gamma^5(39 \Gamma_S^2 \Gamma_D+2 \Gamma_D^3)\\
	&\quad+24 \gamma^4  (30 \Gamma_S^2 \Gamma_D^2+\Gamma_D^4 )+\Gamma_S^2 \Gamma_D^4  (\Gamma_D^2-4\var^2)]\,, \\
	B_8 &= 4t_c^4\Gamma_\phi\{4\gamma^3[\Gamma_S^2(85\Gamma_D^3+44 \Gamma_D \var^2 )+2\Gamma_D^5]\\
	&\quad+\gamma^2[\Gamma_S^2  (78 \Gamma_D^4+88 \Gamma_D^2 \var ^2 )+\Gamma_D^6]
	+\gamma\Gamma_S^2 \Gamma_D^3  (11 \Gamma_D^2+4 \var^2)\}\,, \\
	B_9 &= \Gamma_\phi\{\Gamma_S \Gamma_D\Gamma_\phi\var^2 (6 \gamma +\Gamma_D)
	+ [\gamma\Gamma_\phi^2+2 t_c^2 (4 \gamma +\Gamma_D)][\Gamma_S \Gamma_D (3 \gamma +\Gamma_D)\\
	&\quad+2 t_c^2 (3 \Gamma_S+\Gamma_D)]\}^3.
}	
\end{equation}

\section{Effective model for the Zeno regime}\label{appzeno}

In the regime $\Gamma_D \gg t_c$, the equivalent of the quantum Zeno effect in quantum
transport~\cite{gurvitz98,chen04} occurs. This effect reduces the TQD to an effective DQD described by
equation~\eref{zeno}. The current $I^{\rm Z}$ for the effective DQD is obtained from the known expression for
the current across a DQD~\cite{stoof96}, namely by replacing the rate to the drain $\Gamma_D$
by the effective rate $\Gamma_D^{\rm Z}$ in equation~\eref{effcoup}, and reads
\begin{equation}\label{Idqd}
I^{\rm Z} = \frac{4 \Gamma_S \Gamma_D^{\rm Z} t_c^2}{4 \Gamma_S \varepsilon ^2+\Gamma_S (\Gamma_D^{\rm Z})^2+4 t_c^2 (2 \Gamma_S+\Gamma_D^{\rm Z})}\,.
\end{equation}
This effective current $I^{\rm Z}$ reproduces the average current $I$ for the full TQD system well in
the limit $\Gamma_D \gg t_c$, as we can see from the ratio $I^{\rm Z}/I$ shown in figure~\ref{zenoplot}.
We note that the ratio $I^{\rm Z}/I$ approaches the asymptotic value $1$ in the full Zeno regime,
which is reached for values of $\Gamma_D$ larger than those explored in figure~\ref{zenoplot}. Nevertheless,
for the regime of parameters where we predict DAT, the effect of large $\Gamma_D$ is already partially
taking the system to the Zeno regime. This is noticeable by the suppression of the current in absence
of dephasing, as seen for $\rho_{33}^s$ in figure~\ref{coherence}(a).

\section{Long-distance tunnelling and the effect of pure dephasing on the transport properties of a DQD}\label{appld}

In section~\ref{long} we pointed out that in the regime $\var \gg t_c$ and $\Gamma_D\lesssim t_c$ the long-distance tunnelling
between QD1 and QD3 plays a role in the transport across the TQD. We can understand this effect by first considering the TQD
in absence of dephasing and decoupled from the reservoirs. For large detuning $\var\gg t_c$ it is possible
to eliminate the central dot and describe the system by an effective DQD composed of QD1 and QD3, with the corresponding
Hamiltonian in equation~\eref{lrange} and the effective tunnel coupling $t^{\rm ld}_{13}$ between the outer dots.

We include now the coupling to source and drain reservoirs. Analogous to \ref{appzeno}, the current through the
effective system $I^{\rm ld}$ is obtained by replacing the tunnel coupling $t^{\rm ld}_{13}$ and the effective rates $\Gamma_S^{\rm ld}$
and $\Gamma_D^{\rm ld}$ in the known expression for the current through a DQD, that is
\begin{equation}\label{Idqdld}
I^{\rm ld}=\frac{4 \Gamma_S^{\rm ld} \Gamma_D^{\rm ld} (t^{\rm ld}_{13})^2}{4 \Gamma_S^{\rm ld} (\var^{\rm ld}_{13})^2+\Gamma_S^{\rm ld} (\Gamma_D^{\rm ld})^2
+ 4(t^{\rm ld}_{13})^2(2\Gamma_S^{\rm ld}\!+\!\Gamma_D^{\rm ld})}\,,
\end{equation}
with $\var^{\rm ld}_{13}=\var^{\rm ld}_1-\var^{\rm ld}_3$ and $\var^{\rm ld}_i$ the renormalized on-site energies.

In particular, we can follow the procedure in reference~\cite{paulisch14} to adiabatically eliminate the state
$\ket{2}$, which yields explicitly $t^{\rm ld}_{13}=(\var-\sqrt{\var^2+8t_c^2})/4$ and identical renormalized on-site energies
$\var^{\rm ld}_1 = \var^{\rm ld}_3$. Using these expressions together with $\Gamma_{S/D}^{\rm ld}=\Gamma_{S/D}$
we find that the current $I^{\rm ld}$ reproduces the current $I$ obtained for the full TQD in the regime $\var \gg t_c$
with $\Gamma_D\lesssim t_c$ and $\gamma=0$. This is shown in figure~\ref{zenoplot} in the form of the ratio $I^{\rm ld}/I$,
which tends to unity as the detuning $\var$ increases, thereby demonstrating the validity of the effective description.

The fact that long-distance tunnelling in the full TQD appears only for moderate coupling to the drain
can also be understood with the help of the effective model in~\eref{lrange}: Large values of $\Gamma_D$ induce
a broadening of level $\ket{3}$ that may result in a substantial overlap with the detuned level $\ket{2}$, rendering
the description in terms of an effective DQD invalid. From a dynamic point of view, strong coupling to
the drain takes the effective DQD into a Zeno-like regime, in which the observation of the QD3 tends to freeze the
charge mainly in QD1. This Zeno-like localization becomes dominant once we leave the regime of moderate coupling
$\Gamma_D\sim t_c$, even for large detuning of QD2. It is also interesting that in this regime the Zeno effect reduces
the only coherence participating in the effective DQD, $\rho_{13}^s$, [cf. Fig~\ref{coherence}(b)] and therefore destroys
the long-distance tunnelling in absence of dephasing.

We now turn to the analysis of the effect of dephasing on the transport properties of the effective DQD, i.e.,
we include the Lindblad term $\gen_\phi$ of equation~\eref{super} with the effective rate $\gamma^{\rm ld}$.
As the results are generally valid for a DQD coupled to reservoirs and subjected to pure dephasing we omit
the label \emph{ld} in the following, but apart from this use the same notation as in the main text.

Applying the characteristic polynomial approach~\cite{bruderer13} we obtain the average current
\begin{equation} \label{Idqddeph}
I_{\phi} = \frac{4t_{13}^2\Gamma_S \Gamma_D \Gamma_{\phi}}
{\Gamma_S \Gamma_D (\Gamma_{\phi}^2 + 4\var_{13}^2)+ 4t_{13}^2 \Gamma_{\phi}(2 \Gamma_S+\Gamma_D)}
\end{equation}
and the Fano factor
\begin{equation}\fl\label{Fdqddeph}
F_{\phi} = 1-\frac{8 \Gamma_S\Gamma_D t_{13}^2 \{4 \var_{13}^2[X_{\phi}+\Gamma_D (\Gamma_D -\Gamma_S)]
+\Gamma_{\phi}^2[X_{\phi}+\Gamma_D (3 \Gamma_S + \Gamma_D)+8t_{13}^2]\}}
{[\Gamma_S \Gamma_D (\Gamma_{\phi}^2 + 4\var_{13}^2) + 4t_{13}^2 \Gamma_{\phi}(2 \Gamma_S+\Gamma_D)]^2}\,,
\end{equation}
with $X_{\phi} = 2 \gamma (\Gamma_S + \Gamma_D)$ and $\Gamma_{\phi} = 2 \gamma  +\Gamma_D$.
We notice that $I_\phi$ in equation~\eref{Idqddeph} is equivalent to the current obtained for a DQD coupled to
a bosonic bath within the Born-Markov approximation, if only the contribution of pure dephasing  is considered.
Specifically, the dephasing rate due to an Ohmic bath of phonons is~\cite{ramon04,brandesrep}
$\gamma_{\phi}\propto\alpha\var_{13}^2k_BT/\Delta^2$, with $\Delta=\sqrt{\var_{13}^2 + 4t_{13}^2}$ and $\alpha$
is the strength of the electron-phonon interaction.

In general, the ratio of the current for a DQD with and without dephasing, equations~\eref{Idqddeph} and \eref{Idqdld}
respectively, has the explicit form
\begin{equation}\label{rIDQDdeph}
\frac{I_{\phi}}{I_{0}}=\frac{\Gamma_{\phi}[\Gamma_S(\Gamma_D^2+4\var_{13}^2)
 + 4 t_{13}^2 (2 \Gamma_S+\Gamma_D)]}
{\Gamma_S \Gamma_D (\Gamma_{\phi}^2+4 \var_{13}^2) + 4 t_{13}^2 \Gamma_{\phi}(2 \Gamma_S+\Gamma_D)}\,.
\end{equation}
A detailed analysis of this expression reveals that the current can be either enhanced or suppressed by dephasing,
depending on the relevant parameters of the system.
Note that for the full TQD system we assumed that QD1 and QD3 are in resonance, i.e., $\var_{13}=0$.
In this particular case, equation~\eref{rIDQDdeph} reduces to equation~\eref{suppress} in the main text,
which demonstrates that pure dephasing suppresses the current across a DQD at zero detuning.



\section*{References}

\bibliographystyle{with_titles}

\bibliography{transport}


\end{document}